\documentstyle[preprint,aps]{revtex}
\begin{document}
\tighten
\draft
\preprint{PSU/TH/171; hep-ph/9607282}
\title {
CORRECTIONS OF ORDER $\alpha^6m$ TO $S$-LEVELS OF HELIUM}
\author{Michael I. Eides \cite{emaile}}
\address{Petersburg Nuclear Physics Institute,\\
Gatchina, St.Petersburg 188350, Russia \cite{pmaddr}\\
and\\
Department of Physics, Pennsylvania State University,\\
University Park, Pennsylvania 16802, USA \cite{tmpaddr} }
\author {Howard Grotch \cite{emailg}}
\address{Department of Physics, Pennsylvania State University,\\
University Park, Pennsylvania 16802, USA\medskip}
\date{July 10, 1996}

\maketitle

\begin{abstract}
Several new corrections of order $m\alpha^6$ to the energy levels of 
$S$ states in helium are obtained from radiative corrections to the Breit 
potential and from the polarization insertions in the two photon 
exchange graphs. While individual gauge invariant contributions are comparable 
to current experimental errors, the sum of these corrections is an order 
of magnitude smaller due to mutual cancellations.  
\end{abstract}


Leading logarithmic corrections of order $m\alpha^6\ln\alpha$ to the 
energy levels of helium were recently calculated in \cite{dkmy}.
These corrections are due to the electron-electron interaction and 
they turned out to be phenomenologically relevant for comparison of the 
theory and experiment. Theoretically corrections for helium may be  
directly obtained from the respective contributions to the 
positronium energy levels \cite{fell,kmy} if one knows how to calculate matrix 
elements of the delta function between helium wave functions. 

We have recently obtained some nonlogarithmic corrections of order 
$m\alpha^6$ to the $S$-levels of positronium induced by radiative 
corrections to the Breit potential and  by the polarization 
insertions in the graphs with two-photon exchange\cite{eg}.  Once again 
all such results for positronium may be readily transformed into respective 
corrections to the helium energy levels.  Description of these helium 
corrections is presented below.

Let us start with consideration of the corrections 
to the $S$-levels of helium induced by the diagrams with radiative 
insertions in the graph with the one-photon exchange between 
electrons. As 
is well known this graph leads to the Breit potential. One may easily 
obtain 
the radiatively corrected expression for the Breit potential in the 
form (see, e.g. \cite{gh,eg}\footnote{The expression below differs 
from the 
one in the cited works only by an overall sign, since in the case of 
helium 
we are dealing with a system of two electrons as opposed to the 
electron-positron system in the case of positronium considered in the 
above mentioned papers.})

\begin{equation} \label{breit}
U({\bf p},{\bf r})=\alpha\{\frac{1}{r}
-\pi\frac{1+8m^2f'_1+2f_2}{m^2c^2}\delta^3({\bf r})
+\frac{4\pi p}{m^2c^2}\delta^3({\bf r})
\end{equation}
\[
+\frac{{\bf r}({\bf r}{\bf p}){\bf p}}{2m^2c^2r^3}
+\frac{{\bf p}^2}{2m^2c^2r}
-(3+4f_2)\frac{{\bf s}{\bf l}}{2m^2c^2r^3}
\]
\[
+\frac{(1+f_2)^2}{4m^2c^2}\{s_i,s_j\}(\frac{\delta_{ij}}{r^3}
-3\frac{r_ir_j}{r^5})
-\frac{(1+f_2)^2\pi}{m^2c^2}(\frac{4}{3}{\bf s}^2-2)\delta^3({\bf
r})\},
\]

where $m$ is the electron mass, $\bf p$ is the relative momentum of 
the two electrons, $\bf r$ is their relative position, $f_{1}'$ is the 
slope of the Dirac formfactor, $f_2$ is the Pauli formfactor at zero 
momentum transfer and $p$ is the polarization operator contribution. 
With two-loop accuracy  

\begin{equation} 
f'_1=\frac{e_1}{m^2}\frac{\alpha}{\pi}+\frac{e_2}{m^2}(\frac{\alpha}
{\pi})^2=\frac{\alpha}{3\pi
m^2}(\ln\frac{m}{\lambda}-\frac{3}{8})+\frac{0.469~94}{m^2}
(\frac{\alpha}{\pi})^2,
\end{equation}
\[
f_2=g_1\frac{\alpha}{\pi}+g_2(\frac{\alpha}{\pi})^2
=\frac{\alpha}{2\pi}-0.328~(\frac{\alpha}{\pi})^2,
\]
\[
p=p_1\frac{\alpha}{\pi}+p_2(\frac{\alpha}{\pi})^2
=\frac{\alpha}{15\pi}+\frac{41}{162}(\frac{\alpha}{\pi})^2,
\]

where the two-loop contribution $e_2$ to the slope of the 
Dirac formfactor was calculated numerically in \cite{ab} and 
analytically in 
\cite{bmr}, the two-loop electron magnetic moment $g_2$ was obtained 
in \cite{p,s}, and the value of the two-loop polarization is also well 
known for a long time \cite{ks}.

The radiatively corrected Breit potential above leads to the 
following local 
electron-electron operators which generate corrections of order 
$\alpha^6$ 
to the $S$-levels of helium  

\begin{equation}
V_{F_1}=-0.469\frac{8\alpha^3}{\pi m^2c^2}\delta({\bf r}),
\end{equation}
\[
V_{F_2}=0.328\frac{2\alpha^3}{\pi m^2c^2}\delta({\bf r}),
\]
\[
V_{p2}=\frac{82}{81}\frac{\alpha^3}{\pi m^2c^2}\delta({\bf r}).
\]

Consider now corrections of order $\alpha^6m$ to the energy levels 
of helium generated by the diagrams with high ($\approx m$) 
intermediate 
momenta. These contributions correspond to corrections of order 
$\alpha(Z\alpha)^5m$ and $\alpha(Z\alpha)^5m^2/M$  in case of 
hydrogen.
It is well known that all such corrections are generated by the 
diagrams
with two exchanged photons containing also either polarization 
operator
insertion in one of the exchanged photons or radiative photon 
insertions in
the electron line. External electron lines in the diagrams under
consideration may be safely taken to be on-mass shell with sufficient
accuracy. We are going to consider here the simplest 
gauge-invariant contribution of this kind induced by the one-loop 
polarization operator insertion in the box diagram postponing 
consideration 
of the diagrams with radiative photon insertions in the electron  
line for 
future work.

As we have shown in \cite{eg} contribution to the energy levels 
of positronium of order $m\alpha^6$ induced by the box diagrams with 
one-loop polarization insertion in one of the exchanged photon lines 
is 
given by the expression  

\[
\Delta 
E_{box}=(\frac{\pi^2}{36}-\frac{5}{27})\frac{\alpha^6}{\pi^2n^3}m.
\]

One may easily restore the respective local operator for the 
electron-electron interaction which generates this correction (note 
that 
since this operator is connected with the two-photon exchange it has 
the 
same sign both for the electron-positron and the electron-electron 
interaction)

\[
V_{box}=(\frac{\pi^2}{36}-\frac{5}{27})\frac{8\alpha^3}{\pi m^2}
\delta({\bf r}),
\]

where we have taken into account that for positronium $S$-levels
$|\psi(0)|^2=(m\alpha)^3/(8\pi n^3)$.

Let us collect below final expressions for 
the electron-electron potentials obtained above, using atomic units 
as 
is common in the helium case (technically this means restoring one 
more 
factor of $\alpha$ in the potentials above)

\begin{equation}
V_{F_1}^{hel}=-0.469\frac{8\alpha^4}{\pi m^2}\delta({\bf r}),
\end{equation}
\[
V_{F_2}^{hel}=0.328\frac{2\alpha^4}{\pi m^2}\delta({\bf r}),
\]
\[
V_{p2}^{hel}=\frac{82}{81}\frac{\alpha^4}{\pi m^2}\delta({\bf r}),
\]
\[
V_{box}^{hel}=(\frac{\pi^2}{36}-\frac{5}{27})\frac{8\alpha^4}{\pi 
m^2}
\delta({\bf r}).
\]

Now we may easily obtain numerical values of the corrections to the 
energy 
levels of helium if we use the values of matrix element of the 
operator 
$\pi\delta({\bf r})$ calculated by Drake \cite{drake}. Numerical 
results of 
the calculations are presented in the Table.  Comparing these results 
with 
the experimental data as cited in \cite{dkmy} we see that  while 
individual 
contributions in the Table for $n=2$ are of the same order as the 
experimental errors, the total contribution obtained here is, due to 
mutual 
cancellations, about an order of magnitude smaller than the current 
experimental error.

\acknowledgements

M. Eides is deeply grateful to the colleagues at the Physics 
Department
of the Penn State University for their kind hospitality. This work 
was
supported by the National Science Foundation grant number
PHY-9421408.

\narrowtext

\begin{table}
\caption{Contributions to the Energy Levels}
\begin{tabular}{lrr}    
\        &$1S$ & $2S$ \\
\ $\Delta E$   &  kHz &  kHz
\\ \tableline
$\Delta E_{F_1}$     & $-2374$ & $-193$ 
\\ \tableline 
$\Delta E_{F_2}$     & $414$  &  $34$ 
\\ \tableline 
$\Delta E_{p2}$   & $639$ & $52$ 
\\ \tableline 
$\Delta E_{box}$   & $449$ & $37$ 
\\ 
\end{tabular}
\end{table}

\end{document}